# Transition from conventional ferroelectricity to ion-conduction-like ferroelectricity

Yuanhang Yan and Menghao Wu[*]

School of Physics, Huazhong University of Science and Technology, Hubei 430074, China

**Abstract**

The cross-unitcell long displacements in some recent emergent ferroelectrics have actually challenged the classical definition of ferroelectricity, while the relative explorations are still in the early stage and even controversial. In this paper we provide a general model that gives the picture for the evolution and transition from typical ferroelectricity to long displacement ferroelectricity, which is classified into type-I and type-II. In particular, type-I with two switching modes of different barriers may switch between conventional ferroelectricity and ion-conduction-like ferroelectricity depending on various factors including electric field, boundaries, vacancies, temperature, etc.., which is demonstrated by first-principles calculations on γ-AlOOH and $CuInP_2S_6$ as two paradigmatic cases. Intriguingly, their polarizations are nonlocal since the boundaries also determine the switching mode and polarization direction, which can be different for the same given crystal structure. Such type-I can be evolved from conventional ferroelectricity as the migration barrier across unitcell is reduced, and will behave like type-II at elevated temperature as the conventional part becomes paraelectric. These unconventional behaviors can be applicable to various systems, and many previously unclarified phenomena can be well explained.

## INTRODUCTION

Both ion conduction and ferroelectricity involve ion displacements. However, compared with ion conduction crossing multiple unitcells, the displacements in ferroelectricity are supposed to be within the scale of 1 Å (*1*). Such definition seems to become invalid in some recent emergent ferroelectrics with long ion displacements by moderate barriers(*2*), including many well-known two-dimensional ferroelectrics like $CuInP_2S_6$, $In_2Se_3$, sliding ferroelectrics and layered intercalation ferroelectrics.(*3*) Their unconventional properties like giant quantized polarizations(*4-7*) and violations of Neumann's principle(*8-11*) have been predicted, also blurring the boundary between ferroelectricity and ion conduction.(*12, 13*) However, the explorations for such an abrupt type of ferroelectricity are still in the early stage and even controversial.(*14*) Many intriguing behaviors, including various well states(*15*) (*16*), polarization reversal against the electric field(*17*), etc. are also yet to be clarified.

Here we use a brief well potential model to figure out the continuous evolution from typical ferroelectricity to long displacement ferroelectricity (LDF), with moderate barriers for ions crossing unitcell like ion conductors. The typical double well potential of conventional ferroelectrics is illustrated in Fig. 1(A), where the ion migration crossing unitcell is forbidden by the high barrier. If the migration barrier crossing one unitcell $\Delta E_2$ is only moderately higher compared with ferroelectric switching barrier $\Delta E_1$, as shown in Fig. 1(B), ion conduction may emerge if both barriers can be overcome by an elevated electric field. The two barriers actually correspond to two ferroelectric switching modes of different polarizations, which add up to a polarization quantum $P_0=Z|e||a|/\Omega$ (Z is the valence number, |a| is the lattice constant and $\Omega$ is the unitcell volume) since the displacements add up to a lattice constant. Such bi-mode(*18*) (sometimes called double-path(*19*)) ferroelectricity is denoted as type-I LDF, and the systems include some hydrogen-bonded ferroelectrics(*7, 18, 20*), $CuInP_2S_6$,(*15-17, 21-28*) $HfO_2$(*29*), and $SiO_2$ intercalated by $NH_3$.(*30*) For most ion conductors with non-polar crystals of high symmetry, $\Delta E_1=\Delta E_2$ and periodic potential wells are formed as shown in Fig. 1(C). They may exhibit large quantized(*4*) or fractional quantum ferroelectricity(*8*), denoted as type-II LDF. In some systems like $CuCrX_2$ and $In_2Se_3$ with symmetry breaking perpendicular to the ion conduction channels, the vertical conventional ferroelectricity is coupled with in-plane ion displacement,(*31*) which have been experimentally verified.(*32, 33*) Similarly in typical sliding ferroelectrics like BN bilayer, the vertical ferroelectric switching is achieved by in-plane interlayer sliding of fractional lattice constant.(*34, 35*) For both cases, the ferroelectric switching barriers are exactly the same to their barriers of in-plane ion conduction or interlayer sliding for multiple lattice constants, where continuously ion migration or interlayer sliding lead to oscillation in vertical polarizations, as shown in Fig. 1(D).

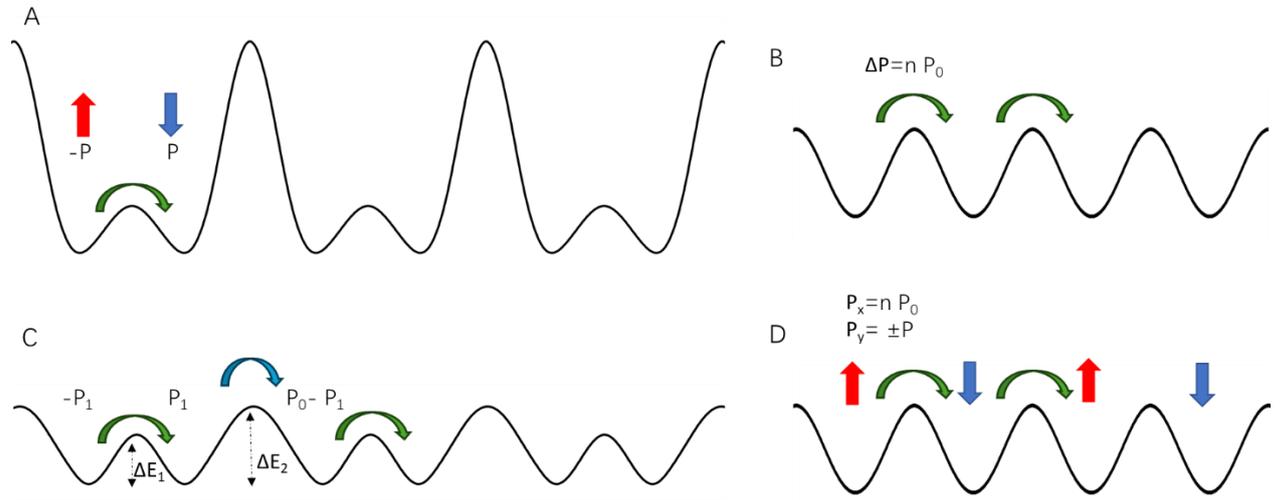

**Fig. 1. Evolution from conventional ferroelectricity to ion-conduction-like ferroelectricity.**

The potential well models for (A) classical ferroelectricity, (B) type-I LDF with two switching modes of different barriers, (C) type-II LDF with crystals of high symmetry where polarization quantum $P_0=Z|e||a|/\Omega$ and n can be integer or fractional number. (D) Some special cases of type-II LDF where the in-plane long displacements are coupled with vertical ferroelectricity.

**RESULTS**

We performed first-principles studies on two paradigmatic cases of type-I LDF, hydrogen-bonded ferroelectrics γ-AlOOH(*18*) and $CuInP_2S_6$. For the case of γ-AlOOH, proton conduction channels are formed by hydrogen-bonded chains. The bi-stable polar states are displayed in Fig. 2(A), and there are two pathways for the transition between them: the protons in state I may either swirl to the right with rotation of -OH, or hop along the hydrogen bonds to the left, both leading to the transformation into state II. However, the former path is driven by an electric field to the right, while the latter path is driven by an electric field to the left, which is inexplicable. Such puzzle for determining the polarization directions of two polar states will be resolved in the following investigations.

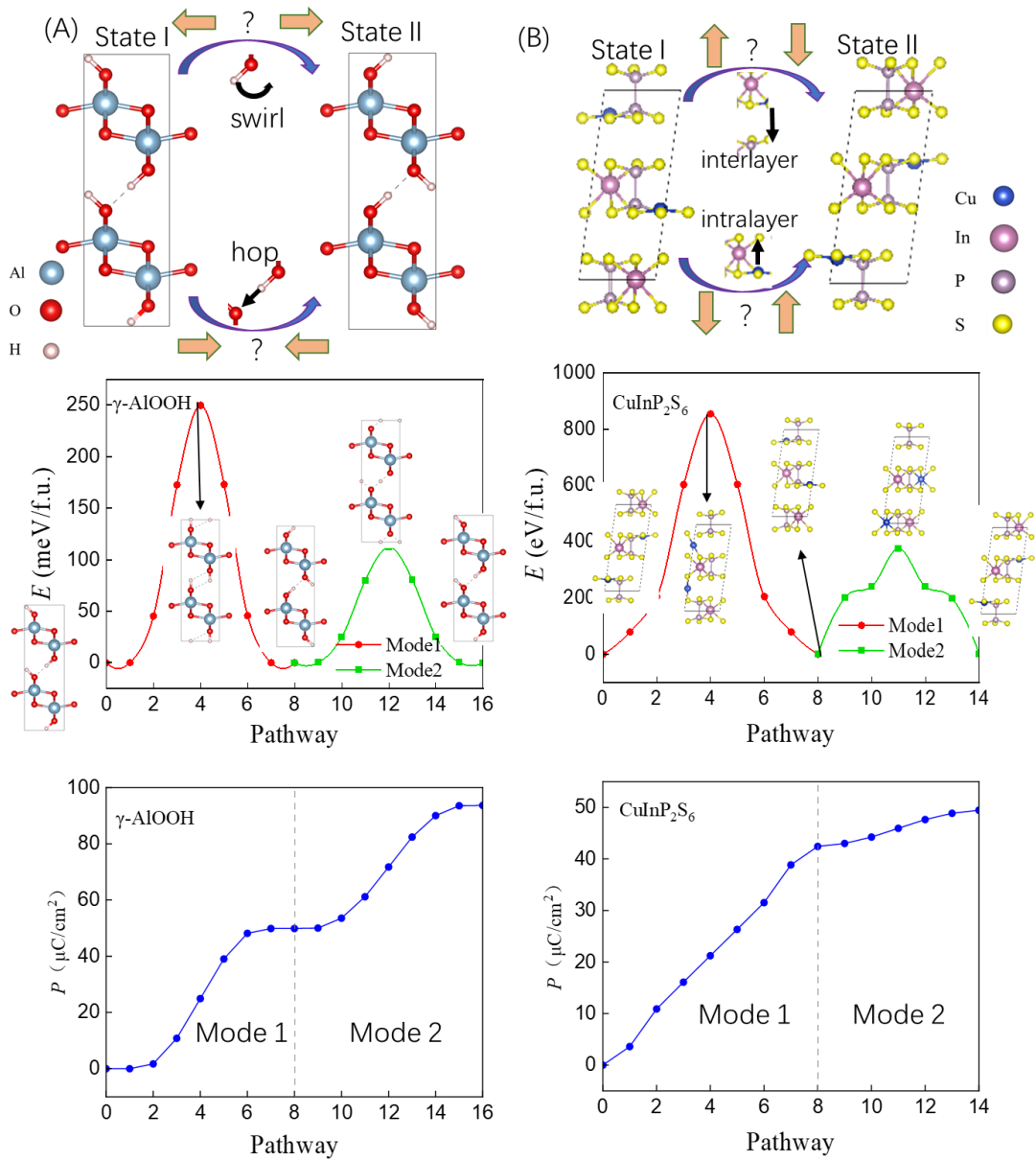

**Fig. 2. Bi-modal ferroelectric switching pathways leading to contradictory definition of polarization directions.**

Ferroelectric bi-stable polar states, two switching modes of different barriers that add up to migration of half lattice constant, and the corresponding evolution of polarization for (A) γ-AlOOH and (B) CuInP$_2$S$_6$.

The proton migration pathway for displacing half lattice constant in Fig. 2(A) is composed of mode 1(swirl) and 2(rotate), with switching barriers of respectively 0.250 and 0.112 eV/f.u.. The evolution of polarization along the pathway reveals that the changes of polarization in mode 1 and 2 are respectively 49.8 and 43.8 μC/cm$^2$, corresponding to $P_1$ and $2P_0-2P_1$ in Fig. 1(B), which add up to $2P_0$ (n=2 since each unitcell contains 4 protons with displacements of half lattice constant |a|/2, where polarization quantum $P_0=|e||a|/\Omega=46.8$ μC/cm$^2$). Similarly for CuInP$_2$S$_6$, there are also two transition pathways between state I and II via the vertical displacements of Cu ions: mode 1 of interlayer migration, and mode 2 of intralayer migration, as shown in Fig. 2(B). Their switching barriers are respectively 0.854 and 0.376 eV/f.u., and the changes in polarization corresponding to $2P_1$ and $P_0-2P_1$ in Fig. 1(B) are respectively 42.4 and 7.0 μC/cm$^2$.

In ideal periodic conditions of crystals, both systems will exhibit conventional ferroelectricity with double-well potential if the electric field can only overcome one of the two barriers ($\Delta E_1$, $\Delta E_2$), and the displacements are confined within unitcell. In this respect, proton hopping will be the favorable switching mode for γ-AlOOH, while the displacements of Cu ions in CuInP$_2$S$_6$ will be confined within layer. If the electric field can overcome both barriers, ions can migrate continuously crossing multiple unitcells along conduction channels like in ion conductors, giving rise to quantized polarizations.

However, if the boundaries are taken into account, the above mentioned displacements may be unfavorable in energy due to the formation of abrupt edges. The slice model of γ-AlOOH with two surfaces is displayed in Fig. 3(A), where state I can be transformed into identical state II with reversed polarization by proton swirling mode to the right. In comparison, proton hopping to the left will lead to the formation of -OH2 at the end of conduction channels (see state II') more than 1 eV higher in energy. Herein it is noteworthy that state II and II' are with exactly the same crystal structure aligned to the same direction (see their unitcells marked in green), but different ion displacements from state I. Therefore the polarizations of state I and II should be respectively towards the left and right.

At the existence of one proton vacancy at one side of each hydrogen-bonded chain, however, proton hopping to the left will give rise to the reversed identical state II', while the formation of state II via swirling to the right is more than 1 eV higher in energy. In this regards, the polarization of state I should be aligned to the right. It seems that polarization of LDF is highly dependent on boundaries, so its domains will be distinct from typical ferroelectric domains with polarization directions simply determined by local structural distortions. The supposed long displacement of protons upon an elevated electric field will actually drive the edge protons out of materials, essentially electrolyzed into hydrogen by high energy cost. If electrolysis by high electric field is not desired, long ion displacements may be realized at existence of either ion vacancies within the conduction channels or ion acceptors at the end.

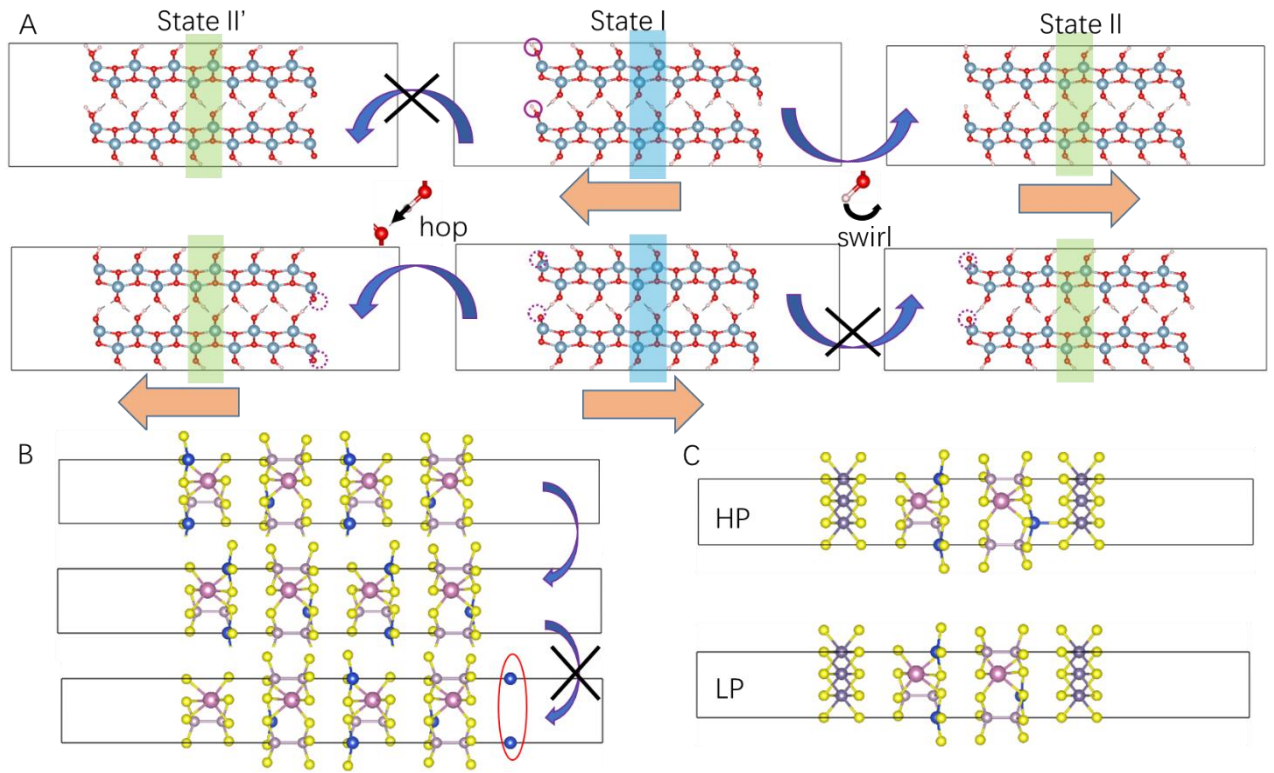

**Fig. 3. Boundary dependence of dominant ferroelectric switching pathways.**

(A) Different switching pathways of γ-AlOOH thin-film determined by different boundaries: proton swirling mode to the right is favorable in the upper panel, which becomes unfavorable compared with hopping mode in the down panel with proton vacancies (marked by the dotted circles) at the edges, and the polarization directions marked by the orange arrows are also different even with –OH groups aligned to the same directions. (B) Long displacements of Cu ions can be unfavorable in isolated few layer $CuInP_2S_6$ that lead to the electrolysis of surface Cu ions (marked by red circle). (C) HP and LP state for few layer $CuInP_2S_6$ covered by $NbS_2$.

Similarly, the long displacements in an isolated few layer $CuInP_2S_6$ will drive the Cu ions of surface layers outside, as shown in Fig. 3(B), which is unfavorable in energy and a high electric field is required for such electrolysis. The energy cost might be reduced when the surface is covered by some strong adsorbent of Cu ions. It is noteworthy that the high polarization (HP) state with Cu ions in vdW gaps under high pressure mentioned in previous studies(*16, 36*) may emerge when the surfaces are covered by 2D materials like NbS$_2$, as shown in Fig. 3(C), which is 0.554 eV/unitcell lower in energy compared with low-polarization (LP) state. Such boundary-dependence may clarify the various different behaviors of CuInP$_2$S$_6$ including quadruple(*16, 27*) and sextuple well states(*15*) observed previously. The reported mysterious behavior of polarization reversal against the electric field(*17*), is also likely a misinterpretation of polarization directions depending on the boundary.

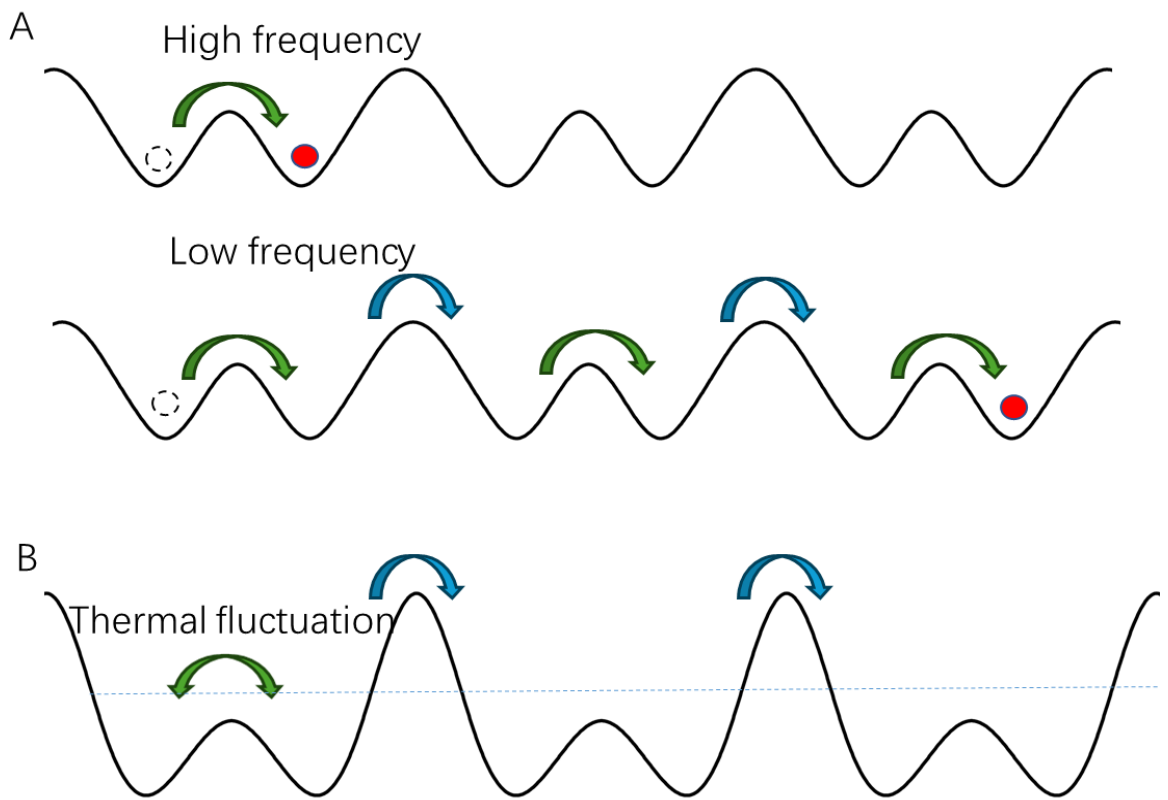

**Fig. 4. Distinct dominant ferroelectric modes upon varying frequency and temperature.** (A) The ion displacements in the hysteresis of LDF will be much smaller in high frequency compared with low frequency electric field. (B) When thermal fluctuation can overcome one of the two barriers in type-I LDF, their conventional part will be paraelectric so they will behave like type-II LDF.

Meanwhile the boundaries will be less important at the existence of ion vacancies, which enable long distance ion migrations even when the ends of conduction channels are blocked.

According to previous studies, in a μm-size single crystal of ion conductors, even 1% ion vacancies may give rise to high switchable polarization over thousands of $\mu C/cm^2$,(*4, 20*) which have been obtained in experiments. (*5, 6*) The frequency of electric field in the hysteresis can also play a role since high frequency does not favor long ion displacements, and long migration is not likely to be achieved within a very small period, as shown in Fig. 4(A). This has been experimentally verified in several proton conductors(*5, 6*) where the giant polarizations are greatly reduced at high frequency electric field.

In addition, ion-conduction-like ferroelectricity can be facilitated by high temperature. When thermal fluctuation can conquer one of the two barriers, the conventional ferroelectricity will diminish due to dipole disorder, while long ion displacements can be driven by an electric field further conquering the other barrier, as shown in Fig. 4(B). Our *ab initio* molecular dynamics simulations in Fig. S1 also reveal that at 700 K, some protons of γ-AlOOH hop to the other side randomly but still confined within the same hydrogen bonds. At 450K, some Cu ions of $CuInP_2S_6$ also become disordered but are still confined within the same $CuInP_2S_6$ layer. In this case, they will behave like type-II LDF with the conventional part in paraelectric state.

## DISCUSSION

In summary, we classified LDF into type-I and type-II, where type-I is key for understanding the transition between conventional ferroelectricity and ion-conduction-like ferroelectricity. They can be evolved from conventional ferroelectricity as the migration barriers across unitcell are reduced, and will behave like type-II at elevated temperature as the conventional part becomes paraelectric. With two switching modes of different barriers, they may exhibit either conventional ferroelectricity or ion-conduction-like ferroelectricity, depending on various factors including electric field, boundaries, vacancies, temperature, etc.. For example, ion-conduction-like ferroelectricity is favored by high electric field with low frequency, while conventional behaviors may emerge upon low electric field with high frequency. Moreover, states of the same crystal structure may give rise to different favorable switching modes polarization directions, which are finally determined by boundaries. Thus such ferroelectricity is nonlocal and the domains will be distinct from typical ferroelectric domains with polarization directions simply determined by local structural distortions. The proposed model may unveil many previous unclarified phenomena, like different reported well states, polarization reversal against the electric field, etc.. Aside from hydrogen-bonded ferroelectrics and CuInP$_2$S$_6$, these unconventional properties can be applicable to various systems including HfO$_2$ and intercalated SiO$_2$, which provide significant scientific and technological opportunities.

## METHODS

Our density functional theory simulations were performed by using the Vienna *ab initio* simulation package (VASP) code(*37, 38*). The generalized gradient approximation in the Perdew-Burke-Ernzerhof (PBE)(*39*) form for the exchange and correlation potential, together with the projector-augmented wave(*40*) method, are adopted. In particular, a van der Waals correction DFT-D3 developed by Grimme et al. (*41*). has been adopted, which could provide a good description of van der Waals interactions. The kinetic energy cutoff is set to be 520 eV, and computed forces on all atoms are less than 0.001 eV/Å after optimization. The Berry phase method is employed to evaluate crystalline polarization(*42*), and the ferroelectric switching pathway is obtained by using the nudged elastic band (NEB) method(*43*).

**Acknowledgements**

**Funding**: This work is supported by the National Natural Science Foundation of China (Nos. 12574263).

**Competing interests**: The authors declare that they have no competing interest.

**Supplementary Materials**

(2000).